\documentstyle[12pt]{article}
\author{S.\ M.\ Troshin and
 N.\ E.\ Tyurin\\
Institute for High Energy Physics\\
Protvino, Moscow Region, 142284 Russia}
\title{\bf Spin content of constituent quarks and
one-spin asymmetries in inclusive processes} \rm \date{}
\begin{document} \maketitle \begin{abstract} We consider
mechanism for one-spin asymmetries observed in inclusive hadron
 production. The main role belongs to the orbital angular momentum of
the quark-antiquark cloud in the internal structure of constituent
quarks. We argue that the origin of the asymmetries in pion
production is a result of retaining of this internal angular orbital
momentum by the perturbative phase of QCD under transition from the
non-perturbative phase. The non-perturbative hadron structure is
based on the results of chiral quark models.

PACS number(s):  11.30.Qc, 12.40.Pp, 13.75.Ni, 13.88.+e
\end{abstract}

\section*{Introduction}
Significant one--spin asymmetries were observed in elastic
 scattering, hyperon and pion production in different reactions
\cite{exp}. A number of models were proposed for qualitative and
quantitative description of the corresponding data.  Despite these
models provide qualitative explanation of the experimental data and
allow in several cases to get quantitative agreement with experiment
it should be noted that there is no single model capable to describe
 all of the existing data. Also new data \cite{newhyp} often pose new
 problems for these models.

In the recent time interest in studying of spin phenomena was shifted
to measurements of spin dependent structure functions and to
extensive interpretations of the data. The substantial experimental
information was obtained by the present time. As it is considered
now, about one third of the proton spin is due to quark spins
\cite{ellis,altar}. It is interesting that calculations of $\eta '$
couplings to vector mesons  also predicted that quarks carry about
one third of spin of vector mesons \cite{band}.  These results could
be interpreted as that a substantial part of hadron spin would be due to
orbital angular momentum of quarks.

The challenging problem is to  relate the  spin
structure of nucleons studied in deep--inelastic scattering with the
 one--spin asymmetries measured in hadron processes.

In this paper we consider a possible origin of asymmetry in pion
production under collision of polarized proton beam with unpolarized
proton target. The experimental data \cite{E704} for such processes were
obtained at rather high energy where one could rely on perturbative
QCD  at high $p_{\perp}$'s.

We will use the scheme which incorporates perturbative and
non--pertur\-ba\-ti\-ve phases of QCD.
We will argue that the orbital angular
momentum of partons inside constituent quarks retained by the
perturbative phase of QCD under transition from non-perturbative
phase leads to significant asymmetries in hadron production with
polarized beam.

 In the nonperturbative regime QCD should provide the two important
phenomena:  confinement and spontaneous breaking of chiral symmetry.
The scales  relevant  to these phenomena are characterized by the
parameters $\Lambda _{QCD}$ and $\Lambda _\chi $, respectively.  The
values of these parameters are $ \Lambda_{QCD}=100-300$ MeV and $
 \Lambda _\chi \simeq   4\pi f_\pi \simeq   1$ GeV, where $f_\pi $
 is  the pion  decay  constant \cite{mnh}.   Chiral $SU(3)_L\times
SU(3)_R$ symmetry is spontaneously  broken  in  the range between
these two scales.  The chiral symmetry breaking results in particular
in generation of quark masses and in appearance of quark condensates.
 Quark masses are comparable with the hadron mass scale.  Therefore
hadron  is often represented as a loosely bounded system of the
constituent quarks.  These observations on the hadron structure lead
to the understanding of several regularities observed in hadron
interactions at large distances. Such a picture also provides
reasonable  values  for the static characteristics of hadrons, for
 instance, their magnetic moments.  Constituent quarks in this
approach are extended objects. They are described by their size and
quark matter distribution.

The  general form of the effective lagrangian relevant for
description of the non--perturbative phase of QCD \cite{gold}
${\cal{L}}_{QCD}\rightarrow {\cal{L}}_{eff}$ includes three terms \[
{\cal{L}}_{eff}={\cal{L}}_\chi +{\cal{L}}_I+{\cal{L}}_C.\label{ef} \]
Here ${\cal{L}}_\chi $ is  responsible for the spontaneous
chiral symmetry breaking and turns on first.  To account for the
constituent quark interaction and confinement the terms ${\cal{L}}_I$
and ${\cal{L}}_C$ are introduced.  The  ${\cal{L}}_I$ and
${\cal{L}}_C$ do not affect the internal structure of constituent
quarks.

However, the structure of hadron  depends on the scale of the process
and is different for different values of $Q^2$. Processes with large
$Q^2$ can resolve partonic structure of constituent quarks and are
described by perturbative QCD. Perturbative  QCD provides well
established calculation methods based on the use of the
${\cal{L}}_{QCD}$.

\section{Structure and spin content of constituent quarks} In this
section we specify the hadron structure and the spin structure of
constituent quarks with account for results obtained in
deep--inelastic scattering experiments.

In the framework of non-perturbative approach we consider a
 hadron as consisting from the valence
constituent quarks located at the central part of a hadron and quark
condensate surrounding this core.  Experimental and theoretical
arguments in favor of such picture were given in \cite{isl,trtu}.
The term ${\cal{L}}_\chi $ provides  masses for quarks and leads to
appearance of the quark condensate.  We consider as a particular form
 of ${\cal{L}}_\chi $ the  Nambu--Jona-Lasinio (NJL) model \cite{njl}
with 6--quark interaction, i. e.  we refer to the version of this
model which takes into account the $U(1)_A$--symmetry breaking term
\cite{jaffe}:  \begin{eqnarray} {\cal L}_\chi  & = & \bar{\psi
}(i\gamma \cdot \partial -\hat{m})\psi \, +
\,\sum_{a=0}^8\frac{1}{2}\,G\,[\,(\,\bar{\psi } \lambda _a \psi
\,)^2+ (\,\bar{\psi }i\lambda _a\gamma _5\psi \,)^2\,]+\nonumber\\ &
& K [\det \bar{\psi }_i(1-\gamma _5)\psi_j + \det \bar{\psi
}_i(1+\gamma _5)\psi_j ], \label{L} \end{eqnarray} where quark field
$\psi $ has three colors $(N_c=3)$ and three flavors $(N_f=3)$ and
matrix $\hat{m}=diag (m_u,\,m_d,\,m_s)$ is composed from the current
quark masses.  Eq. (\ref{L}) may be considered as a minimal effective
lagrangian which reflects some of the  basic  properties  of
nonperturbative QCD. The last term in Eq. (\ref{L}) obeys the chiral
$SU(3)_L\times SU(3)_R$ invariance, but it breaks the unwanted
$U(1)_A$-- symmetry. The four--fermion lagrangian of the NJL model
reveals this symmetry in the $N_f\geq 3$ case.  The first two terms
represent the well--known NJL  lagrangian.  These terms ensure the
dynamical breaking of the $SU(3)_L\times SU(3)_R$ chiral symmetry
when the coupling constant  $G$  is large enough. It has been shown
that the chiral symmetry is broken dynamically and quark acquires a
mass when the coupling constant $G$ is beyond its critical value.
The lagrangian (\ref{L}) in addition to the $4$--fermion interaction
of the original NJL  model includes $6$--fermion $U(1)_A$--breaking
term.  The constituent quark masses have been calculated in
\cite{jaffe}:  \begin{equation}
m_U  = m_u-2G\langle 0|\bar u u|0\rangle-2K\langle 0|\bar d
d|0\rangle \langle 0|\bar s s|0\rangle \end{equation}

In this approach massive  quarks appear  as quasiparticles, i.e. as
current quarks and the surrounding  clouds  of quark--antiquark pairs
which consist of a mixture of quarks of different flavors.
Therefore besides its mass, quark acquires an internal structure  and
a finite size.  Quark radii are determined by the radii of  the
clouds surrounding it.  We assume that the strong interaction radius
 of  quark  $Q$  is determined by its Compton wavelength:
\begin{equation} r_Q=\xi /m_Q,  \label{rq} \end{equation} where
 constant $\xi$ is universal for different  flavors. Quark
 formfactor $F_Q(q)$ is taken in the dipole form, viz
\begin{equation} F_Q(q)\simeq (1+\xi^2{\vec{q}}^{\,2}/m_Q^2)^{-2}
\label{ff} \end{equation} and the corresponding quark matter
distribution $d_Q(b)$ is of the form \cite{trtu}:  \begin{equation}
d_Q(b)\propto \exp(-{m_Qb}/{\xi}). \label{bf} \end{equation} Quantum
numbers of the constituent quarks are the same as the quantum numbers
of current quarks due to conservation of the corresponding currents in
QCD.  The only exception is the flavor--singlet, axial--vector
current, its $Q^2$--dependence is due to axial anomaly which arises
 under quantization.  Axial anomaly gives contribution in the spin
content of the constituent quark as it was discussed in
 \cite{altar,fri,elwa,frin}.  In particular, it was demonstrated
 that constituent quark picture of a hadron with account for anomaly
contribution is consistent with the results for the proton spin
structure function $g_1(x)$ obtained in deep--inelastic scattering.

 It is useful to note that in addition to $u$ and $d$ quarks
constituent quark ($U$, for example) contains pairs of
strange quarks, and the
ratio of scalar density matrix elements \begin{equation} {2\langle U|
 \bar s s|U\rangle}/ {\langle U|\bar u u+\bar d d|U\rangle}
 \label{str} \end{equation} is about 0.15 in the NJL model with axial
 $U(1)$ breaking \cite{stein1}. It should be noted, however, that the
 following inequalities are valid for different mechanisms
\begin{equation} \langle U| \bar u u|U\rangle \gg \langle U| \bar  d
d|U\rangle,\, \langle U| \bar  s s|U\rangle.\label{qc} \end{equation}

 The picture of hadron consisting from constituent quarks can be
 applicable at  moderate momentum transfers, while interactions at
 high momentum transfers would resolve internal structure of
 constituent quarks and they are to be represented as clusters of
 non-interacting partons in this kinematical region.

 In the framework of the NJL model transition to partonic picture is
 related to the need of a momentum cutoff $\Lambda=\Lambda_\chi\simeq
 1$ GeV.  We adopt the point  that the need for such cutoff is an
 effective implementation of the short distance behavior in QCD
\cite{jaffe}.

 Thus, loosely speaking we should consider three different regions
 for hadron structure depending on the typical scale of interaction:
 interactions with small momentum transfers ($Q<\Lambda_{QCD}$) do
  not resolve internal structure of hadrons, interactions with medium
 momentum transfers ($\Lambda_{QCD}<Q<\Lambda_{\chi}$) see hadrons as
 consisting from constituent quarks and interactions with high
 momentum transfers ($Q>\Lambda_{\chi}$) resolve the partonic
 structure of constituent quarks. Of course, this separation is
 approximate and real picture is definitely more complicated.

 In the framework of the NJL model the partonic structure of
 constituent quarks was defined in \cite{jaffe}. The parton content
 $\omega_{q/Q}(x)$ of constituent quark $Q$ as it was shown
is determined
 by the imaginary part of the virtual antiquark--quark scattering
amplitudes $t_{1\bar q/Q}(s,\mu^2)$ and $t_{2\bar q/Q}(s,\mu^2)$ and
can be written as follows:  \[
\omega_{q/Q}(x)=\pi^2\int_{s_{th}}^\infty \frac{ds}{(2\pi)^3}
\int_{-\infty}^{{\mu_{max}}^2}d\mu^2 \mbox{Im}[t_{1\bar
q/Q}(s,\mu^2)+xt_{2\bar q/Q}(s,\mu^2)], \] where $s$ is the squared
 center of mass energy for   the quark--antiquark scattering and
 $\mu^2$ is the virtual quark squared mass; $\omega_{q/Q}(x)$ is the
 distribution of flavor $q$ quarks in the constituent quark $Q$.

Now we will address a complicated problem of the spin structure of
constituent quark.  The measurements of spin--dependent structure
function $g_1(x)$ triggered the discussion of the role of axial
anomaly in the nucleon spin.
In
the framework of perturbative theory
 it was argued that axial anomaly in QCD
effectively reduces the total spin carried by  quarks \cite{efrem}.

On the other hand the contribution of axial anomaly could have a
non-perturbative origin. For example, in the NJL--model the 6-quark
fermion operator in Eq. (\ref{L}) simulates the effect of gluon
operator \[\frac{\alpha_s}{2\pi}G^a_{\mu\nu}\tilde G^{\mu\nu}_a,\]
where $G_{\mu\nu}$ is the gluon field tensor in QCD.  Account for
axial anomaly in the framework of chiral quark models results in
 compensation of the valence quark helicity by  helicities of quarks
 from the cloud in the structure of constituent quark. The specific
 non-perturbative mechanism of such compensation is different in
 different approaches \cite{jaffe,fri,elwa,frin},  e.g. in
  \cite{elwa} the modification of the axial U(1) charge of
  constituent quark is generated by the interaction of current quarks
with flavor singlet field $\varphi^0$.  Forward matrix elements of
 the currents $A^j_{\mu 5}$ between the constituent quarks $Q^i$ are
 written then as follows \begin{equation} \langle Q^i|A^j_{\mu
 5}|Q^i\rangle= \frac{1}{2}(\delta^i_j-\frac{2}{3}c)s^i_{\mu},
 \label{sp} \end{equation} where $s^i_\mu$ is the spin vector and the
 constant $c$ determines the derivative coupling between quarks and
 field $\varphi^0$.  Eq. (\ref{sp}) shows that  constituent quark of
 any flavor contains a sea of polarized current quarks of the all
 other
  flavors.  The case of $c=1/2$ corresponds to  complete compensation
 of current quark spins.  The similar picture was developed by
 Fritzsch \cite{fri}. On this ground we can conclude that significant
 part of the spin of constituent quark should be associated with the
 orbital angular momentum of quarks inside this constituent
 quark, i.e. the cloud quarks should rotate coherently inside
 constituent quark.
  We consider effective lagrangian approach where gluon degrees
 of freedom are overintegrated
  and therefore we are not going to discuss subtle
 questions on the principal possibility of separation between
  the orbital angular momentum and gluon contribution in QCD
  (cf. \cite{kiss}).

  The important question concerns the origin of this orbital angular
  momentum. It is useful to address an analogy between hadron physics
  and superconductivity, in particular, anisotropic generalization
  of the theory of superconductivity which seems to match well with
  the above picture for constituent quark.  Indeed, it was shown
  \cite{anders,gaitan} that pairing
  correlations have axial symmetry around the anisotropy direction $
  \hat {\vec l}$ which acts as the local $z$ axis. Because of this
  anisotropy there are particle currents induced by pairing
   correlations. The corresponding calculations \cite{anders}
  indicate that  particle at the origin is surrounded by a cloud of
  correlated particles that rotate around it with the axis of
  rotation $\hat {\vec l}$.  The value of intrinsic orbital angular
  momentum $L_0$ is determined by the density of particles $\rho $,
  the gap amplitude $\Delta_0$ and by the Fermi energy $E_f$:  \[
  L_0=\frac{1}{2}(\rho-C_0)\simeq
  \rho\left(\frac{\Delta_0}{E_f}\right)^2 \] It is clear that there
  is a direct analogy between the above picture  and that of
 constituent quark.  Axis of
 anisotropy $\hat {\vec l}$ is determined by the polarization vector
 of  valence quark located at the origin of constituent quark. The
  orbital angular momentum $\vec L$ lies along $\hat {\vec l}$ and
  its value is proportional to quark density.

  Thus, the spin of constituent quarks $J_{zU}$ is determined by the
  following sum \begin{equation} J_{zU}=1/2=J_{zu_v}+J_{z\{\bar q
  q\}}+\langle L_{z\{\bar q q\}}\rangle.  \label{bal} \end{equation}
 The value of the orbital momentum contribution into the spin of
 constituent quark can be estimated with account for new experimental
 results from deep--inelastic scattering \cite{altar} indicating
 that quarks carry one third of proton spin, i.e.  \[ (\Delta u
 +\Delta d +\Delta s)_p\simeq 1/3, \] and taking into account  the
 relation between contributions of current quarks into  proton spin
and corresponding contributions of current quarks into the spin of
constituent quarks and contributions of constituent quarks into the
proton spin \begin{equation} (\Delta u +\Delta d +\Delta s)_p =
(\Delta U+\Delta D) (\Delta u +\Delta d +\Delta s)_U.\label{qsp}
\end{equation} Indeed, if we adopt $SU(6)$ model ($\Delta U+\Delta
D=1$) then we should conclude  that \[ J_{zu_v}+J_{z\{\bar q
   q\}}\simeq 1/6\] and from Eq. (\ref{bal}) \[ \langle L_{z\{\bar q
  q\}}\rangle\simeq 1/3,\] i. e. about 2/3 of the $U$-quark
   spin is
due to the orbital angular momenta of $u$, $d$ and $s$ quarks inside
$U$-quark.  Index $z$ will be dropped henceforth.

 We argue that the existence of this orbital angular momentum, i.e.
 orbital motion of quark matter inside constituent quark, is the
 origin of the observed asymmetries in inclusive production at
  moderate and high transverse momenta.  Indeed, since the
 constituent quark has small size, asymmetry associated with internal
 structure of this quark will be significant at
 $p_{\perp}>\Lambda_\chi\simeq 1$ GeV where interactions at small
 distances  give noticeable contribution.

 The orbital motion of current quarks means that they have intrinsic
 transverse momenta. Estimation of the mean value of this momenta
  from the relation \[ \langle  L_{\{\bar q q\}}\rangle=r_Q\langle
k_\perp\rangle \] with $\langle L_{\{\bar q q\}}\rangle\simeq 1/3$
 and $r_Q= 1/3$--$1/6$ fm provides the values of $200$--$400$ MeV
which are in agreement with experimental values. Note, that
these estimations correspond to $\xi\simeq 1/3$ in Eq. (\ref{rq}).

 It should be noted that at high $p_\perp$ we will have a parton
 picture for constituent quark as a cluster of non-interacting quarks
 which however should naturally preserve  their orbital
momenta of the preceding non-perturbative phase of QCD, i.e. the
 orbital angular momentum will be retained in perturbative phase of
QCD.

 \section{Model of hadron production and one-spin asymmetries}
  Consider now mechanism of hadron production based on the above
 picture of hadron structure. We will study the hadron processes of
 the  type \[ h_1^\uparrow +h_2\rightarrow h_3 +X \] with polarized
 beam or target.

   The picture of hadron consisting from constituent quarks embedded
 into quark condensate implies that overlapping and interaction of
peripheral clouds   occur at the first stage of hadron interaction.
As a result massive
virtual quarks appear in the overlapping region and  some effective
field is generated.
 Constituent quarks  located in the central part of hadron are
supposed to scatter in a quasi-independent way by the effective
 field.
 In the above picture generation of
the effective field is related with the term ${\cal L}$ and
formation of the final hadrons should be described by the
 term ${\cal L}_C$ in the Lagrangian ${\cal L}$.

  Inclusive production of hadron $h_3$  results from
 recombination of the constituent quark
 (low $p_{\perp}$'s, soft interactions) and from the
 excitation of this constituent quark, its decay and subsequent
 fragmentation in the hadron $h_3$. The latter process is determined
 by the distances smaller than constituent quark radius and is
 associated therefore with hard interactions (high $p_{\perp}$'s).
   Thus,
 we adopt the two--component picture of hadron production which
  incorporates
  the non-perturbative and perturbative QCD phases.

Now we write down explicit formulas for corresponding inclusive
cross--sections. The following expressions  were obtained in
\cite{tmf} and take into account unitarity in the direct channel of
reaction. They have the form \begin{equation}
\frac{d\sigma^{\uparrow,\downarrow}}{d\xi}= 8\pi\int_0^\infty
bdb\frac{I^{\uparrow,\downarrow}(s,b,\xi)} {|1-iU(s,b)|^2},\label{un}
\end{equation} where $b$ is the impact parameter. Here  function
$U(s,b)$ is the generalized reaction matrix (helicity non-flip one)
which is determined by  dynamics of the elastic reaction \[
h_1+h_2\rightarrow h_1+h_2. \] The elastic scattering amplitude $F$
  (at the moment we consider spinless case for simplicity) is related
  to the function $U$ by the  equation:  \begin{equation} F=U+iUDF,
\label{6} \end{equation} which we write here in the operator form.
This equation allows one to obey unitarity provided inequality
  $ \mbox{Im}\,U(s,b)\geq 0\,$  is fulfilled.

 In accordance with the quasi-independence of valence quarks we
represent the basic dynamical  quantity in  the form  of  product
\cite{trtu}:  \begin{equation} U(s,b)\,=\, \prod^N_{i=1}\, \langle
f_{Q_i}(s,b) \rangle \label{prd} \end{equation} in the impact
parameter representation, $N=n_{h_1}+n_{h_2}$ is the total number of
constituent quarks in the initial hadrons.  Factors $\langle
f_{Q}(s,b)\rangle$ correspond to the individual quark scattering
 amplitude smeared over transverse position of $Q$ inside hadron
  $h_1$ and over fraction of longitudinal momentum of the initial
  hadron carried by quark $Q$.

The functions $I^{\uparrow,\downarrow}(s,b,\xi)$ are related to the
functions $U_n (s,b,\xi,\{\xi _{n-1}\})$ which are the multiparticle
analogs of the $U(s,b)$ and are determined  by  dynamics
of the processes \[ h_1^{\uparrow,\downarrow}+h_2\rightarrow
h_3+X_{n-1}.  \] The kinematical variables $\xi$ ($x$ and $p_\perp$,
 for example) describe the state of the produced particle $h_3$ and
 the set of variables $\{\xi_{n-1}\}$ describe the system $X_{n-1}$
 of $n-1$ particles. Arrows $\uparrow$ and $\downarrow$ denote
 corresponding  direction of  transverse spin of the polarized
 initial  particle.

Expressions for the functions $I^{\uparrow,\downarrow}$ are the
following \cite{tmf}:  \begin{equation}
 I^{\uparrow,\downarrow}(s,b,\xi)= \sum_{n\geq
3,\lambda_2,\lambda_{X_n}} n \int d\Gamma_n'
|U^{\uparrow,\downarrow}_{n,\lambda_2, \lambda_{X_n}} (s,b,\xi,
\{\xi_{n-1}\}|^2, \end{equation} where $X_n=h_3+X_{n-1}$.  In the
above formulas  $d\Gamma_n'$ is the element of  $n-1$-particle phase
space volume.

 We introduce the two functions $I_+$ and $I_-$:  \begin{equation}
I_{\pm}(s,b,\xi)=I^\uparrow(s,b,\xi)\pm I^\downarrow(s,b,\xi),
\end{equation} where  $I_+(s,b,\xi)$ corresponds to unpolarized case.
The following sum rule takes place for the function $I_+(s,b,\xi)$:
\begin{equation} \int I_+(s,b,\xi)d\xi=\bar n(s,b)\mbox{Im}
U(s,b),\label{sr} \end{equation} where $\bar n(s,b)$ is the mean
multiplicity of secondary particles in the impact parameter
representation.

Asymmetry $A_N$  defined as the  ratio \[ A_N=
\{\frac{d\sigma^\uparrow}{d\xi}-\frac{d\sigma^\downarrow}{d\xi}\}/
\{\frac{d\sigma^\uparrow}{d\xi}+\frac{d\sigma^\downarrow}{d\xi}\} \]
can be expressed in terms of the functions $I_{\pm}$ and $U$:
\begin{equation} A_N=\frac{\int_0^\infty bdb
I_-(s,b,\xi)/|1-iU(s,b)|^2} {\int_0^\infty bdb
I_+(s,b,\xi)/|1-iU(s,b)|^2}.\label{xnn} \end{equation}

Using relations between transversely polarized states
$|\uparrow,\downarrow\rangle$ and helicity states $|\pm\rangle$, viz
\begin{equation}
|\uparrow,\downarrow\rangle=(|+\rangle\pm|-\rangle)/\sqrt{2}
\end{equation} one can write down expressions for  $I_+$
and $I_-$ through the helicity functions $U_{\{\lambda_i\}}$:
\begin{eqnarray} I_+(s,b,\xi) & = & \sum_{n,\lambda_1,
\lambda_2,\lambda_{X_n}} n \int d\Gamma_n' |U_{n,\lambda_1,\lambda_2,
\lambda_{X_n}} (s,b,\xi, \{\xi_{n-1}\})|^2, \nonumber\\ I_-(s,b,\xi)
 & = & 2\sum_{n, \lambda_2,\lambda_{X_n}} n \int d\Gamma_n'
\mbox{Im}[ U_{n,+,\lambda_1,\lambda_2, \lambda_{X_n}} (s,b,\xi,
\{\xi_{n-1}\})\times \nonumber\\ & & U^*_{n,-,\lambda_1,\lambda_2,
 \lambda_{X_n}} (s,b,\xi, \{\xi_{n-1}\})]. \end{eqnarray}

Taking into account Eq. (\ref{sr}), quasi-independence of the
 constituent quarks and assumption on hadron production as a result
of interaction of the corresponding constituent quark with the
effective field we adopt the following expressions for the functions
$I_+(s,b,\xi)$ and $I_-(s,b,\xi)$:  \begin{equation} I_\pm
(s,b,\xi)=\bar n(s,b)\mbox{Im}[\prod_{i=1}^{N-1} \langle
f_{Q_i}(s,b)\rangle\langle
\varphi^\pm_{h_3/\tilde{Q}}(s,b,\xi)\rangle], \label{prod}
\end{equation} where quark $\tilde Q$ is the leading quark in the
process of $h_3$ production, for example, $\tilde Q=U$ for
$h_3=\pi^+$ and $\tilde Q=D$ for $h_3=\pi^-$. The functions
$\langle\varphi^\pm_{h_3/\tilde{Q}}(s,b,\xi)\rangle$ describe the
  $h_3$ production as a result of interaction of the
constituent quark $\tilde Q$ with the effective field.

The central point of the model is a connection  of the one-spin
asymmetries in inclusive production with the orbital angular momentum
of current quarks inside the constituent quark. This orbital momentum
will affect the hadron production only at small distances where
internal structure of constituent quark could be probed. Thus, the
function $\langle\varphi^-_{h_3/\tilde{Q}}(s,b,\xi)\rangle$ will be
sensitive to interactions at small distances only, i.e. it will be
 determined by the hard processes which can be described in the
framework of perturbative QCD, but with account for the internal
orbital momentum of partons.  This function can be written as the
convolution integral:  \begin{equation}
\langle\varphi^-_{h_3/\tilde{Q}}\rangle= \langle\varphi^-_{\tilde
q/\tilde{Q}}\rangle\otimes D_{h_3/{\tilde q}}, \end{equation} where
$D_{h_3/{\tilde q}}$ is the fragmentation function which is supposed
to be spin-independent. Owing to inequalities (\ref{qc}) the leading
contribution is given by the  quark $\tilde q$ of the same flavor as
$\tilde Q$. Due to isospin invariance $D_{\pi^+/u}=D_{\pi^-/d}$.
Fragmentation functions are almost completely unknown quantities in
QCD.  We consider these functions to be spin independent and
  the asymmetries are related here to the internal structure of
  constituent quarks.  Note that the fragmentation functions might
have a non-trivial spin dependence as it was discussed in
\cite{cols}.

Spin-independent function $\langle\varphi^+_{h_3/\tilde{Q}}\rangle$
gets contribution both from soft processes, where constituent quark
interacts with the effective field as a whole (hadron $h_3$ arises in
this case as a result of recombination of $\tilde Q$ with the virtual
quarks) and from hard interactions associated with partonic structure
 of the constituent quark.  Respectively, the function
$\langle\varphi^+_{h_3/{\tilde Q}}\rangle$ contains the two terms,
 viz \begin{equation} \langle\varphi^+_{h_3/\tilde{Q}}\rangle=
\langle\varphi^+_{\tilde{Q}}\rangle\otimes D_{h_3/{\tilde Q}}+
\langle\varphi^+_{\tilde q/\tilde{Q}}\rangle\otimes D_{h_3/{\tilde
q}}, \end{equation} where first term corresponds to soft and second
 one -- to hard, spin-independent interactions. Since the second term
 in this formula and the function $\langle \varphi^-_{h_3/{\tilde
 Q}}\rangle$ are determined by the internal structure of constituent
quark, then we can assume that their $x$-dependence is determined by
the structure function of constituent quark $\omega_{\tilde q/\tilde
Q}(x)$:  \begin{equation} \langle\varphi^\pm_{\tilde q/\tilde
Q}\rangle\otimes D_{h_3/\tilde q} \propto \omega_{\tilde q/\tilde
Q}(x)\label{w} \end{equation} while \begin{equation}
\langle\varphi^\pm_{\tilde Q}\rangle\otimes D_{h_3/\tilde Q} \propto
\omega_{\tilde Q/h_1}(x), \label{ww} \end{equation} where
$\omega_{\tilde Q/h_1}(x)$ is the $x$-distribution of the constituent
quark $\tilde Q$ in the hadron $h_1$.

It is an important question at this point what is the effect of
nonzero orbital momentum of quarks and consequently their internal
transverse momenta $\langle k_{\perp}\rangle$ inside the constituent
quark.  It leads to a certain shift of transverse momenta and  on
   the basis of
Fourier transformation we suppose that this effect results in the
phase factor $\exp{[ i\langle k_{\perp\tilde q/\tilde Q}\rangle
r_{\tilde Q}]}$ since the $b$-dependence of the functions
$\langle\varphi^{\pm}_{\tilde q/\tilde Q}\rangle\otimes D_{h_3/\tilde
q}$ is determined by  formfactor of the constituent quark $\tilde{Q}$
(cf. Eqs. (\ref{ff}), (\ref{bf})).  On this ground we adopt the
following relation between the spin-dependent function
$\langle\varphi^-_{\tilde q/\tilde Q}\rangle$ and the
 spin-independent function $\langle\varphi^+_{\tilde q/\tilde
 Q}\rangle$:  \begin{equation} \langle\varphi^-_{\tilde q/\tilde
 Q}\rangle\otimes D_{h_3/\tilde q} \simeq \exp{[ i\langle
 k_{\perp\tilde q/\tilde Q}\rangle r_{\tilde Q}]}
\langle\varphi^+_{\tilde q/\tilde Q}\rangle\otimes D_{h_3/\tilde
q}.\label{*} \end{equation} Taking into account that the  orbital
 angular momentum of $\tilde{q}$ quarks in the constituent
quark $\tilde Q$ is
proportional to its polarization we can rewrite exponential factor of
Eq.(\ref{*}) in the form \[ \exp{[ i\langle k_{\perp\tilde q/\tilde
 Q}\rangle r_{\tilde Q}]} = \exp{[ i\langle L_{\tilde q/\tilde
Q}\rangle ]} \simeq \exp{[i{\cal{P}}_{\tilde Q}\langle L_{\{q\bar
q\}}\rangle]}, \] where $\langle k_{\perp\tilde q/\tilde Q}\rangle$
 and $\langle L_{\tilde q/\tilde Q}\rangle$ are the mean transverse
momenta and orbital momenta respectively of quark $\tilde q$ inside
quark $\tilde Q$.  The sign and value of the latter are determined by
the polarization ${\cal{P}}_{\tilde Q}$ of the constituent quark
$\tilde Q$ inside the hadron $h_1$ and the mean orbital momenta of
cloud quarks $\langle L_{\{q\bar q\}}\rangle$.

Taking into account the above relations, we  represent the asymmetry
  $A_N$ in the form:  \begin{equation} A_N(s,x,p_{\perp})= \frac{
\sin[{\cal{P}}_{\tilde Q}\langle L_{\{\bar q q\}}\rangle]
\omega_{{\tilde q}/\tilde Q}(x)\phi_{hard} (s,p_{\perp})}
{\omega_{{\tilde Q}/h_1}(x)\phi_{soft} (s,p_{\perp}) +\omega_{{\tilde
q}/\tilde Q}(x)\phi_{hard} (s,p_{\perp})},\label{an} \end{equation}
where the  function $\phi _{hard}$ is determined by the interactions
at small distances and reflects the structure of  constituent quarks
while $\phi _{soft}$ is associated with the soft interactions and
determined by a non-perturbative structure of  hadron consisting from
its constituent quarks. The explicit forms of these functions are
determined by the integrals entering Eq. (\ref{xnn}).  We can rewrite
 Eq. (\ref{an}) in more general form \begin{equation}
A_N(s,x,p_{\perp})= \sin[{\cal{P}}_{\tilde Q}\langle L_{\{\bar q
q\}}\rangle] \frac{d\sigma_{hard}}{d\xi}/
\{\frac{d\sigma_{soft}}{d\xi}+ \frac{d\sigma_{hard}}{d\xi}\},
\label{ann} \end{equation} which is appropriate for numerical
analysis.

\section{Numerical analysis} Prior to discussions of the experimental
data it should be noted that the described mechanism is to be
expected to work at high enough energies and transverse momenta when
the structure of constituent quark can be resolved.

Explicit form of $A_N$ is determined by the specific
parameterizations of quark and hadron formfactors, corresponding
structure functions, mean multiplicity and several other
distributions. These parameterizations imply rather large freedom and
will unfortunately obscure the main features of the proposed
mechanism.

Therefore, as a first step to numerical analysis of the data, it
seems reasonable to use phenomenological parameterizations of
inclusive cross-sections which can be matched with the above model as
an input to obtain  asymmetry $A_N$.  Indeed, for such purposes we
should consider a two-component parameterization of inclusive
cross-sections which includes soft and hard contributions \[
\frac{d\sigma}{d\xi}= \frac{d\sigma_{soft}}{d\xi}+
\frac{d\sigma_{hard}}{d\xi}.  \] The parameterization of such type
was used under analysis of the experimental data for cross-sections
of the processes:  \[ p+p\rightarrow \pi^\pm+X \] at different
 energies \cite{par}. Due to this we consider  asymmetries $A_N$ in
the processes with polarized initial proton:  \[
p_{\uparrow}+p\rightarrow\pi^\pm+X.  \] Asymmetry for the process \[
p_{\uparrow}+p\rightarrow\pi^0+X \] will be obtained using the
following relation valid in a parton model:  \begin{equation}
A_N(\pi^0)= \frac{A_N(\pi^+)\frac{d\sigma}{d\xi}(\pi^+)+
A_N(\pi^-)\frac{d\sigma}{d\xi}(\pi^-)}
{\frac{d\sigma}{d\xi}(\pi^+)+\frac{d\sigma}{d\xi}(\pi^-)}.
\label{pio} \end{equation} Eq. (\ref{pio}) follows also from  the
isospin relations for one-particle inclusive productions \cite{soff}.
 Now to get values of asymmetries we should fix ${\cal{P}}_{\tilde
Q}$ and $\langle L_{\{\bar qq\}} \rangle$ for $U$-quarks
($\pi^+$-production) and for $D$-quarks ($\pi^-$-production).  For
 polarization of the constituent quarks we  use $SU(6)$ values
 ${\cal{P}}_U=2/3$ and ${\cal{P}}_D=-1/3$.  Orbital angular momentum
was estimated in sec. 2, therefore we take $\langle L_{\{\bar q
q\}}\rangle=1/3$.  The explicit form for $d\sigma/d\xi$ and values of
the parameters therein we borrowed from \cite{par}. It has typical
two-component behavior:  \begin{eqnarray} \frac{d\sigma}{d\xi} & = &
A\exp[-B\sqrt{p_{\perp}^2+m_0^2}\,]
/(1+e^{D(x_\perp-x_0)})+\nonumber\\ & &
C(1-x_\perp)^m(p_{\perp}^4+M^4)^{-n/4},\label{ds} \end{eqnarray}
where $x\simeq 0$ and  $A$ and $m_0$ both have a weak energy
dependence.  The first term in Eq. (\ref{ds}) has typical form of
soft contribution and will be identified with $d\sigma_{soft}/d\xi$
and the second one, decreasing as a power of $p_\perp$, is typical
for hard contribution and is to be identified with
$d\sigma_{hard}/d\xi$.  Thus, we have all parameters fixed and can
evaluate now the asymmetries $A_N$ at different energies.  At high
energies the experimental data for the process
$p_{\uparrow}+p\rightarrow \pi^0+X$ are available at $P_L=200$ GeV/c.
 Comparison of the
calculations for $A_N$ with the data and predictions for asymmetries
 in the processes of $\pi^\pm$ and $\pi^0$ production at this energy
 as well as at $P_L=70$, $800$ GeV/c and $\sqrt{s}=500$, 2000 GeV are
given in Figs. 1--4.  $A_N$ has a weak energy dependence and
gets significant values starting from $p_{\perp}\simeq 1$ GeV/c. As
it is seen from Fig. 3 asymmetry $A_N$ for the process
$p_{\uparrow}+p\rightarrow \pi^0+X$ predicted by the model is
systematically higher that the experimental data. This fact could
confirm conclusion that hadron wave function deviates from $SU(6)$
  model as it was claimed in \cite{artru}.  To check this statement
we calculated the above asymmetries  with
${\cal{P}}_U=-{\cal{P}}_D={\cal{P}}_p$.  The results are presented in
 Figs.  5-8. As it is clearly seen the agreement with the
experimental data is better than for the case of $SU(6)$ model.  It
is also evident that asymmetries in the production of charged pions
are significantly higher than under the neutral pion
production.  This indicate that studies of the charged pion
production would reveal significant asymmetries which are diluted in
the case of neutral pion production. The corresponding values
 will allow to get conclusion on the  mean orbital angular
momenta of quark matter inside the constituent quarks.

\section{Summary and discussion}

First, we would like to summarize the main points of the considered
model:  \begin{itemize} \item asymmetry reflects internal structure
of the constituent quarks and is proportional to the orbital angular
momentum of current quarks inside the constituent quark; \item sign
of asymmetry and its value are proportional to polarization of the
 constituent quark inside the polarized initial hadron, in the simplest
 case this polarization is determined by the $SU(6)$-symmetry.
\end{itemize}

  We have not considered here quantitative description of the
$x$--dependence of $A_N$, first of all, because it includes
rather large freedom under the choice of explicit parameterization.
Indeed, the realistic $x$--dependencies of the functions
$\langle\varphi^\pm_{\tilde q/\tilde Q}\rangle\otimes D_{h_3/\tilde
q}$ and $\langle\varphi^\pm_{\tilde Q}\rangle\otimes D_{h_3/\tilde
Q}$ are definitely more complicated than those indicated in Eqs.
(\ref{w}) and (\ref{ww}). We should consider corresponding
convolution integrals of the structure functions
(of constituent quarks, current quarks inside constituent quarks) and
fragmentation functions. Eqs. (\ref{w}) and (\ref{ww}) show only
characteristic parts of these dependencies.   The
realistic choice of the corresponding parameterizations of the
structure and fragmentation functions as well as choice of
 the $x$-dependence of constituent quark polarization would allow to
get  description of the $x$--dependence of asymmetries.

The model predicts significant one-spin asymmetries at high
$p_{\perp}$ values.  At first sight it looks like contradiction since
the model itself was inspired by QCD where we should expect helicity
 conservation in hard region due to the chiral $SU(3)_L\times
SU(3)_R$ symmetry. Indeed, asymmetry $A_N$ results from interference
between the two helicity functions $U_{+\lambda_2,\lambda_n}$ and
$U_{-\lambda_2,\lambda_n}$ and therefore from the helicity conservation
rule for exclusive processes \cite{brlp} \[
\lambda_1+\lambda_2=\lambda_n\] we have to expect $A_N=0$ at high
$p_\perp$'s. However, the helicity conservation rule at hadron level
is not a direct consequence of the chiral symmetry of the perturbative
phase of QCD.  In addition to helicity conservation at quark level
it assumes
that only S-state of quarks contributes to a hadron wave function. This
statement was disputed in the  work of Ralston and Pire \cite{rlst}
where it was demonstrated that helicity may not be conserved at
 hadron level. The origin of such effect is due to a nonzero orbital
 angular
momentum component of the hadron wave function.

In our model the orbital angular momentum plays a role in the wave
function of constituent quarks.  The two helicity functions
$U_{\pm\lambda_2,\lambda_n}$  in the impact parameter representation
gain different phase factors due to internal transverse momentum of
partons related with their coherent rotation inside constituent
quark.  It results in interference between these two functions and
leads  to significant values of $A_N$.  This mechanism is not at
all at variance with QCD.

Number of recent papers demonstrated that large asymmetries observed
in inclusive processes do not contradict to QCD.  Different
mechanisms were proposed as a source of the asymmetries:  higher
twist effects \cite{efremt}, correlation of $k_\perp$ and spin in
structure \cite{sivs} and fragmentation \cite{cols,artru} functions,
rotation of valence quarks inside a hadron \cite{boros}.  Significant
role in the above references belongs to orbital angular momentum of
 the constituents inside a hadron.  It is worth to note here that this
idea could be traced back to the model of rotating hadronic matter
proposed by Chou and Yang \cite{yang}.

As it was already argued the main role belongs to the orbital angular
momentum of current quarks inside the constituent quark while
constituent quarks themselves have very slow (if at all) orbital
motion and may be described approximately by $S$-state of the
 hadron wave function.  The observed $p_{\perp}$-behavior of
asymmetries in inclusive processes seems to confirm such conclusions.
The significant asymmetries appear to show up beyond $p_{\perp}>1$
GeV/c, i.e.  the scale where internal structure of a constituent
quark can be probed.

The proposed mechanism, in principle, is appropriate for description
of hyperon polarization, in particular, its $p_{\perp}$-dependence.
We can assume that constituent quark $\tilde Q$ gets polarization due
to multiple scattering in effective field by analogy with mechanism
proposed in  \cite{szwed}. Then polarization of $\Lambda$-hyperon
will be proportional to constituent quark polarization and
polarization of $s$-quark inside constituent quark $\tilde Q$.  The
latter one can be related to the significant $s$-quark polarization
measured in deep-inelastic scattering.  Eq. (\ref{str}) also
indicates that  constituent quarks have significant strangeness
content.  The $p_{\perp}$-dependence of
polarization ${\cal{P}}_\Lambda$ will be related to the specific
behavior of soft and hard contributions to inclusive cross-section of
 $\Lambda$-production, but the general trends are expected to be the
same as in $p_{\perp}$-behavior of asymmetry in $\pi^-$-production.
We should also expect that behavior of asymmetry $A_N$ in
$\Lambda$-production  will be similar to the corresponding behavior
of polarization ${\cal{ P}}_\Lambda$ and this fact seems  find
confirmation in the data at $P_L=200$ GeV/c \cite{bravar}.

\subsection*{Acknowledgements} We would like to thank  N. Akchurin,
M. Anselmino, A. Bravar, W.-D. Nowak and V. Petrov for useful
 discussions.

\newpage \normalsize \begin{center} \bf Figure captions \end{center}
\rm \bf Fig. 1. \rm Asymmetry $A_N$ ($SU(6)$ model) in the process
$p_{\uparrow}+p\rightarrow \pi^++X$ (positive values) and in the
process $p_{\uparrow}+p\rightarrow \pi^-+X$ (negative values) at
$P_L=70$ GeV/c (dashed curve), $P_L=200$ GeV/c (solid curve) and at
$P_L=800$ GeV/c (dashed-dotted curve).\\[2ex] \bf Fig. 2. \rm
Asymmetry $A_N$ ($SU(6)$ model) in the process
$p_{\uparrow}+p\rightarrow \pi^0+X$ at $P_L=70$ GeV/c (dashed curve),
$P_L=200$ GeV/c (solid curve) and at $P_L=800$ GeV/c (dashed-dotted
curve). Experimental data from \cite{E704}.\\[2ex] \bf Fig. 3. \rm
Asymmetry $A_N$ ($SU(6)$ model) in the process
$p_{\uparrow}+p\rightarrow \pi^++X$ (positive values) and in the
process $p_{\uparrow}+p\rightarrow \pi^-+X$ (negative values) at
$\sqrt{s}=500$ GeV (dashed curve) and at $\sqrt{s}=2000$ GeV
(dashed-dotted curve).\\[2ex] \bf Fig. 4. \rm Asymmetry $A_N$
($SU(6)$ model) in the process $p_{\uparrow}+p\rightarrow \pi^0+X$ at
$\sqrt{s}=500$ GeV (dashed curve) and at $\sqrt{s}=2000$ GeV
(dashed-dotted curve).\\[2ex] \bf Fig. 5. \rm Asymmetry $A_N$ for the
case ${\cal{P}}_U=-{\cal{P}}_D={\cal{P}}_p$ in the process
$p_{\uparrow}+p\rightarrow \pi^++X$ (positive values) and in the
process $p_{\uparrow}+p\rightarrow \pi^-+X$ (negative values) at
$P_L=70$ GeV/c (dashed curve), $P_L=200$ GeV/c (solid curve) and at
$P_L=800$ GeV/c (dashed-dotted curve).\\[2ex] \bf Fig. 6. \rm
Asymmetry $A_N$ for the case ${\cal{P}}_U=-{\cal{P}}_D={\cal{P}}_p$
in the process $p_{\uparrow}+p\rightarrow \pi^0+X$ at $P_L=70$ GeV/c
(dashed curve), $P_L=200$ GeV/c (solid curve) and at $P_L=800$ GeV/c
(dashed-dotted curve). Experimental data from \cite{E704}.\\[2ex] \bf
Fig. 7. \rm Asymmetry $A_N$ for the case
${\cal{P}}_U=-{\cal{P}}_D={\cal{P}}_p$ in the process
$p_{\uparrow}+p\rightarrow \pi^++X$ (positive values) and in the
process $p_{\uparrow}+p\rightarrow \pi^-+X$ (negative values) at
$\sqrt{s}=500$ GeV (dashed curve) and at $\sqrt{s}=2000$ GeV
(dashed-dotted curve).\\[2ex] \bf Fig. 8. \rm Asymmetry $A_N$ for the
case ${\cal{P}}_U=-{\cal{P}}_D={\cal{P}}_p$ in the process
$p_{\uparrow}+p\rightarrow \pi^0+X$ at $\sqrt{s}=500$ GeV (dashed
curve) and at $\sqrt{s}=2000$ GeV (dashed-dotted curve).


\small \begin{thebibliography}{99} \bibitem{exp} A. D. Krisch,
Summary talk given at 11th Int. Symp. on High Energy Spin Physics,
Bloomington, 15-23 September 1994.  \bibitem{newhyp} J. Duryea et
al., Phys. Rev. Lett. \bf 67 \rm, 1193 (1991).\\ A. Morelos et al.,
Phys. Rev. Lett. \bf 71 \rm, 2172 (1993).  \bibitem{ellis} J. Ellis
and M. Karliner, CERN-TH.7022/93 (1993) and 7394/94 (1994).
\bibitem{altar} G. Altarelli and G. Ridolfi, CERN-TH.7415/94 (1994).
\bibitem{band} M. Bander, Phys. Rev. \bf D44, \rm 3695 (1991).
\bibitem{E704} D. Adams et al. (FNAL E704 Collaboration), IHEP 94-88
 (1994).  \bibitem{mnh} A.  Manohar and H.  Georgi, Nucl. Phys. \bf
B234\rm, 139 (1984).  \bibitem{gold} T. Goldman and R. W.  Haymaker,
Phys. Rev.  \bf D24\rm, 724 (1981).  \bibitem{isl} R.  D.  Ball,
Intern.  Journ. of Mod. Phys. \bf A 5\rm, 4391 (1990);\\ M. M.
Islam, Zeit.  Phys. C \bf  53\rm, 253 (1992); Foundation of Phys. \bf
24\rm, 419 (1994); \bibitem{trtu} S. M. Troshin and N. E. Tyurin,
 Phys.  Rev. \bf D49 \rm, 4427 (1994).  \bibitem{njl} Y.  Nambu and
 G.  Jona-Lasinio, Phys.  Rev. \bf 122\rm, 345 (1961).
\bibitem{jaffe} V.  Bernard, R. L. Jaffe and U.--G.  Meissner, Nucl.
Phys. \bf B308\rm, 753 (1988);\\ S.  Klimt, M. Lutz, V.  Vogl and W.
Weise, Nucl. Phys. \bf A516\rm. 429 (1990);\\ T.  Hatsuda and T.
Kunihiro, Nucl.  Phys. \bf  B387\rm, 715 (1992); \bibitem{fri} H.
Fritzsch, Phys. Lett. \bf A5\rm, 625 (1990); Phys. Lett. \bf B256\rm,
75 (1991); CERN-TH.7079/93 (1993).  \bibitem{elwa} U. Ellwanger and
B. Stech, Phys. Lett. \bf B241\rm, 449 (1990); Z. Phys. C \bf 49\rm,
683 (1991).  \bibitem{frin} S. Forte, Phys. Lett. \bf B224\rm, 189
(1989);\\ R. L. Jaffe and H. J. Lipkin, Phys. Lett. \bf B266\rm, 458
(1991);\\ A. E. Dorokhov and N. I. Kochelev, Phys. Lett. \bf B259\rm,
335 (1991);\\ K. Steininger and W. Weise, Phys. Rev. \bf D48\rm, 1433
(1993).  \bibitem{stein1} K. Steininger and W. Weise, Phys. Lett. \bf
B329\rm, 169 (1994).  \bibitem{efrem} A. V. Efremov and O. V.
Teryaev, Dubna E2-88-287 (1988);\\ G. Altarelli and G. G. Ross, Phys.
 Lett. \bf B212\rm, 391 (1988);\\ R. D. Carlitz, J. C. Coliins and A.
 H. Mueller, Phys. Lett. \bf B214\rm, 229 (1988).  \bibitem{kiss} R.
L. Jaffe and A. Manohar, Nucl. Phys. \bf B337, \rm  509 (1990);\\ A.
V. Kisselev and V. A. Petrov, Theor. Math. Phys. \bf 91\rm, 490
(1992).  \bibitem{anders} P. W. Anderson and P. Morel, Phys. Rev. \bf
123\rm, 1911 (1961).  \bibitem{gaitan} F. Gaitan, Annals of Phys. \bf
235\rm, 390 (1994).  \bibitem{tmf} S. M.  Troshin and N.  E.  Tyurin,
Teor. Mat. Fiz. \bf 28\rm, 139 (1976); Z. Phys. C \bf 45\rm, 171
(1989).  \bibitem{soff} J. Soffer and D. Wray, WIS-72/31, 1972.
\bibitem{cols} J. C. Collins, Nucl. Phys. \bf B396\rm, 161 (1993);\\
J. C. Collins, S. F. Heppelman and G. A. Ladinsky, PSU/TH/101, 1993.
\bibitem{par} E. W. Beier et al., Phys. Rev. \bf D18\rm, 2235 (1978).
\bibitem{brlp} S. J. Brodsky and G. P. Lepage, Phys. Rev. \bf D24\rm,
2848 (1981).  \bibitem{rlst} J. P. Ralston and B. Pire, Preprint
Kansas 5-15-92, 1992 \bibitem{artru} X. Artru, J. Czy\.{z}ewski and
H. Yabuki, LYCEN/9423, TPJU 12/94 1994.  \bibitem{efremt} A. V.
Efremov and O. V. Teryaev, Sov. J.  Nucl. Phys. \bf 36\rm, 140
(1982);\\ J. Qiu and G. Sterman, Nucl. Phys. \bf B378\rm, 52
(1992);\\ S. J. Brodsky, P. Hoyer, A. H. Mueller and W.-K. Tang,
Nucl. Phys. \bf B369\rm, 519 (1992).  \bibitem{sivs} D. Sivers, Phys.
Rev. \bf D41\rm, 83 (1990), ibid. \bf D43\rm,  261 (1991);\\ M.
Anselmino, M. E. Boglione and F. Murgia, DFTT47/94, INFNCA-TH-94-21,
1994; Talk given at 11th Int. Symp. on High Energy Spin Physics,
Bloomington, 15-23 September 1994.  \bibitem{boros} C. Boros, Liang
Zuo-tang and Meng Ta-chung, Phys. Rev. Lett \bf 70\rm, 1751 (1993);\\
M. Doncheski, MAD/PH/778, 1993.  \bibitem{szwed} J. Szwed, Phys.
Lett. \bf B105\rm, 403 (1981).  \bibitem{yang} T. T. Chou and C. N.
Yang, Nucl. Phys. \bf B107, \rm 1 (1976), Proc. Sixth Int. Symp.
Polar. Phenom. in Nucl. Phys., Osaka, 1985 J. Phys. Soc. Jpn. \bf
 55\rm, 53 (1986).  \bibitem{bravar} A. Bravar (FNAL E-704
Collaboration), Private communication.  \end{thebibliography}
\end{document}